\documentclass[conference]{IEEEtran}

\IEEEoverridecommandlockouts
\usepackage{cite}

\usepackage{amsmath,amsthm, amssymb, amsfonts}
\usepackage{array}
\usepackage{multicol}
\usepackage{orcidlink}
\usepackage{amssymb}
\usepackage{algorithm}
\usepackage{algpseudocode}
\usepackage{graphicx}
\usepackage{textcomp}
\usepackage{comment} 
\usepackage{xcolor}
\def\BibTeX{{\rm B\kern-.05em{\sc i\kern-.025em b}\kern-.08em
    T\kern-.1667em\lower.7ex\hbox{E}\kern-.125emX}}
\usepackage{glossaries}

\newglossaryentry{realnumbers}{
  name={\ensuremath{\mathbb{R}}},
  description={the set of real numbers},
  sort={real numbers}
}

\newglossaryentry{RWTH}{
  type=\acronymtype,
  name={RWTH},
  description={Rheinisch-Westfälische Technische Hochschule Aachen},
  first={Rheinisch-Westfälische Technische Hochschule Aachen (RWTH)},
  sort={Rheinisch-Westfälische Technische Hochschule Aachen}
}
\newacronym{SOMC}{SOMC}{Service-Oriented Model-Based Control}
\newacronym{SOA}{SOA}{Service-Oriented Architecture}
\newacronym{BO}{BO}{Bayesian Optimization}
\newacronym{RMS}{RMS}{Root Mean Square}
\newacronym{ASOA}{ASOA}{Automotive Service-Oriented Architecture}
\newacronym{TOPSIS}{TOPSIS}{Technique for Order Preference by Similarity to Ideal Solution}
\newacronym{QoS}{QoS}{Quality of Service}
\newacronym{SOC}{SOC}{Service Oriented Computing}
\newacronym{MPC}{MPC}{Model Predictive Controller}
\newacronym{AI}{AI}{Artificial Intelligence}
\newacronym{IoT}{IoT}{Internet of Things }
\newacronym{ROS}{ROS}{Robot Operating System}
\usepackage{tabularx}
\usepackage{import}
\usepackage{newtxtext,newtxmath}
\usepackage{multirow}
\theoremstyle{definition}
\newtheorem{definition}{Definition}
\newtheorem*{remark}{Remark}
\usepackage{csquotes}
\usepackage{graphicx}
\usepackage{subcaption}
\usepackage{mathtools}
\usepackage{bm} 

\usepackage{tikz}
\usetikzlibrary{arrows.meta,positioning,calc}
\usepackage{ps}
\preprinttrue
\publishedfalse
\input{copyright.tex}
\hypersetup{
    colorlinks = false,
    pdfborder = {0 0 0}
}
\begin{document}
\title{Computation–Accuracy Trade-Off in Service-Oriented Model-Based Control
{\footnotesize \textsuperscript{}}
\thanks{This research is supported by the Deutsche Forschungsgemeinschaft
(German Research Foundation) with the grant number 468483200.}
\thanks{
            $^{1}$ The authors are with the Department of Aerospace Engineering, University of the Bundeswehr Munich, Germany, {\texttt{hazem.ibrahim@unibw.de, julius.beerwerth@unibw.de, bassam.alrifaee@unibw.de}}}
\thanks{
            $^{2}$ The author is with the Institute of Automatic Control, RWTH Aachen University, Germany, {\texttt{l.doerschel@irt.rwth-aachen.de}}}}

\author{Hazem Ibrahim$^{1}$\,\orcidlink{0009-0007-1911-4760}, Julius Beerwerth$^{1}$\,\orcidlink{0000-0001-6167-9692}~\IEEEmembership{Student~Members,~IEEE}\\
Lorenz Dörschel$^{2}$\,\orcidlink{0000-0001-9566-6371}, ~\IEEEmembership{Member,~IEEE}, and Bassam Alrifaee$^{1}$\,\orcidlink{0000-0002-5982-021X},~\IEEEmembership{Senior Member,~IEEE}
}

\maketitle
\thispagestyle{IEEEtitlepagestyle}
\pagestyle{plain}

\begin{abstract}
Representing a control system as a \gls{SOA}—referred to as \gls{SOMC}—enables runtime-flexible composition of control loop elements. This paper presents a framework that optimizes the computation–accuracy trade-off by formulating service orchestration as an A$^\star$ search problem, complemented by Contextual \gls{BO} to tune the multi-objective cost weights. A vehicle longitudinal-velocity control case study demonstrates online, performance-driven reconfiguration of the control architecture. We show that our framework not only combines control and software structure but also considers the real-time requirements of the control system during performance optimization.
\end{abstract}


\section{Introduction}
\label{sect:intro}
\glspl{SOA} decompose software into smaller, independent units called services that can be distributed across different hardware systems \cite{b18}. 
Services follow principles such as reusability and composability, enabling them to form service compositions\cite{b18}. 
The process of creating and adapting these compositions is called orchestration, managed by a central entity known as the orchestrator. 
Within \gls{SOA}, orchestration plays a central role by dynamically connecting services to fulfill system-level functions\cite{b18}.

Control system architectures traditionally rely on fixed component interactions\cite{b19}. Searching for the same concept of \gls{SOA} in control theory, several methods aim to create runtime-flexible control systems. Switching adaptive control can detect changes and toggle between pre-defined controllers which offers some runtime flexibility but remain limited\cite{b28}. For overactuated systems, control allocation redistributes control values to compensate for actuator failures \cite{b29}. However, these techniques predetermine all components and their interactions during the design phase of the control system, limiting flexibility to a set of predefined options. An example of adapting control system architectures is the dynamic updating of control systems for evolving self-adaptive systems introduced in \cite{b30}. This approach focuses on identifying changes in the objective and then constructing control systems using components that offer dynamic control or deployment methods. Another approach is the Metacontrol for the \gls{ROS} (MROS) framework. It is a model-based system for real-time adaptation of robot control system architectures using ROS. It combines domain-specific languages to model various architectural options. MROS also implements the MAPE-K cycle (Monitor, Analyze, Plan, Execute, Knowledge) and meta-control frameworks for dynamic adaptation using an ontology-based approach\cite{b31}.

The previous solutions focus solely on control, without considering the software structure. Applying the \gls{SOA} concept to control system architectures differs from its conventional use in web or enterprise domains, mainly due to the real-time and performance requirements of control systems. We extend this concept through our \gls{SOMC} framework\cite{b13}, where each control loop element is modeled as a service and integrated at runtime by a central orchestrator. Our previous work \cite{b13}, based on the \gls{ASOA} by Kampmann et al. \cite{kampmann2022asoa}, demonstrated how runtime integration and updatability can be achieved in embedded systems \cite{b41}. However, the earlier architecture \cite{b13} relied on predefined scenarios in the form of state machines for orchestration and did not explicitly consider the control accuracy or the computational cost of the selected service composition.  As a result, the orchestration process could not guarantee optimal trade-offs between control accuracy and computation time.

In the domain of \gls{SOA}, multiple approaches have been proposed to use graph-based methods to solve the optimal service composition problem. For example, \cite{b10} implemented a graph-based orchestration method using web services and an \gls{AI} graph-planning algorithm to select optimal service paths. W. Bi et al. \cite{b11} employed k-dimensional trees and nearest-neighbor search for cloud service selection, demonstrating fast and effective service composition. H. Alhosaini et al. \cite{b20} proposed a hierarchical method based on skyline services. Skyline services are a type of service that acts as a gateway or intermediary between client applications and other backend services. This method precomputes Pareto-optimal skylines to optimize \gls{QoS}-driven service composition, improving efficiency through precomputation and caching. In mobile edge computing, J. Wu et al. \cite{b15} introduced M3C, an optimization method for micro-service composition that minimizes latency and energy use, offering practical deployment benefits.
Beside this method, \cite{b23} developed a top-k QoS-optimal service composition method for \gls{IoT} systems, leveraging service dependency graphs and dynamic programming to reduce search complexity and improve performance. Similarly, \cite{b24} proposed a heuristic polynomial-time graph search algorithm for web service composition. \cite{b25} introduced the Pre-joined Service Network, which uses graph databases to retrieve optimal service compositions efficiently. In \cite{b40}, Kampmann et al optimized resource allocation in \gls{SOA} to reduce power consumption and execution time. Searching for optimal performance in control, L. P. Fröhlich et al. \cite{b12} optimized lap time using contextual \gls{BO} for tuning of \gls{MPC}. Most of these methods are applied to optimize performance in web technologies, which do not include real-time requirements.

In this paper, we move from \gls{ASOA} to \gls{ROS} 2 and introduce a graph-based orchestration framework that uses the A$^\star$ search algorithm \cite{b16} to evaluate and select the optimal service composition with respect to a multi-objective cost function. To automatically determine the weighting between computation time and control accuracy, we employ Contextual \gls{BO} \cite{b12}, which learns the optimal cost weights from experimental data. This framework allows the orchestrator to adapt its decisions when control objectives or operating conditions change, enabling online, performance-driven reconfiguration of the control architecture while considering its real-time requirements.  

We organize the remainder of the paper as follows. We present the preliminaries for \gls{SOMC} in Sec. \ref{sect:prelim}. After that, Sec. \ref{sect:methimpl} discusses our graph-based framework consisting of two phases: service graph creation and graph-based orchestration. Sec. \ref{sect:evaluation} shows the results of applying our framework to a vehicle longitudinal velocity control example. Finally, we conclude in Sec.\ref{sect:conclusion} and provide an outlook on future work.

\section{Preliminaries}
\label{sect:prelim}
In this section, we briefly introduce some definitions within \gls{SOMC}.
\begin{definition}[Control System]
The control system is defined by sensors, filters, controllers, actuators, and process forming a control loop.
\end{definition}
\begin{definition}[Control System Services]
Control system services are representative forms of the control system elements that produce or consume information, as shown in Figure~\ref{fig:basic_control_loop}, such as sensors, filters, controllers, models, processes, and actuators. Let \( \mathbb{S} = \{s_1, \ldots, s_n\} \) denote the set of services in a control system, where \( |\mathbb{S}| = n \in \mathbb{N}^{>0} \) and $n$ is the total number of available control system services. Each service \( s_i \in \mathbb{S} \), \(\forall \;1 \leq i \leq n\), possesses:
\begin{itemize}
    \item a set of guarantees \( G_i = \{g^1_i, \ldots, g^m_i\} \), interpreted as outputs representing the information provided by the service, where \( |G_i| = m \in \mathbb{N}^{>0} \) and $m$ is the total number of guarantees of the service $s_i$. For example, in Figure~\ref{fig:basic_control_loop}, the filter service  guarantees to provide the estimated system states $\tau_x$. 
    \item a set of requirements \( R_i = \{r^1_i, \ldots, r^q_i\} \), interpreted as inputs representing the information required by the service, where \( |R_i| = q \in \mathbb{N}^{\geq 0} \) and $q$ is the total number of requirements of the service $s_i$. For example, in Figure~\ref{fig:basic_control_loop}, the filter service requires measured values  $\tau_y$, a control signal $\tau_{u}$, and a model $\tau_{model}$.
\end{itemize}
The set \(\mathbb{S} \) can be categorized into six subsets
according to their roles in the control system such that:
\[
S_{\text{sensor}}, \; S_{\text{filter}}, \; S_{\text{controller}}, \; S_{\text{actuator}}, \; S_{\text{model}}, \; S_{\text{process}} \subseteq \mathbb{S}
\]
\end{definition}
\begin{definition}[Orchestrator in a Service-Oriented Control System]

The \emph{orchestrator} \( \mathcal{O} \) is a supervisory entity defined as a function $\mathcal{P}$:
\[
\mathcal{O} : \mathcal{P}(\mathbb{S}) \rightarrow \mathcal{C},
\]
that, given the set of available services \( \mathbb{S} \), produces a configuration
\[
\mathcal{C} = \{ (s_a, s_b) \mid s_a, s_b \in \mathbb{S},\; G_a \cap R_b \neq \emptyset \},
\]
representing the directed interconnections between services whose guarantees satisfy the requirements of downstream services.
\end{definition}

The orchestrator selects one representative service from each subset such that:
\[
S_{\text{sel}} = \{ s_{\text{sensor}}, s_{\text{filter}}, s_{\text{controller}}, s_{\text{actuator}}, s_{\text{model}}, s_{\text{process}} \}
\]
So the control system formulated by $S_{\text{sel}}$ will be in the following form:
\[
s_{\text{sensor}} \rightarrow s_{\text{filter}} \rightarrow s_{\text{controller}} \rightarrow s_{\text{actuator}} \rightarrow s_{\text{process}}
\]
(optionally involving \( s_{\text{model}} \)), such that all information dependencies are satisfied:
\[
\forall s_i \in S_{\text{sel}}, \; R_i \subseteq \bigcup_{s_j \in S_{\text{sel}}, j \neq i} G_j.
\]

\section{Methodology}
\label{sect:methimpl}
This section provides a step-by-step explanation of our framework in \gls{SOMC}. Sec. \ref{sect:graphcreation} presents the definition of the service graph, which is created from the available control system services. Sec.\ref{sect:orchworkflow} explains of our orchestration method that uses A$^\star$ and contextual \gls{BO} to determine the optimal service composition.

\subsection{Service Graph Definition}
\label{sect:graphcreation}
In computer science, graphs are often used to model communication networks \cite{b8}. A graph is composed of a set of nodes and a set of edges that connect the nodes. Depending on the nature of the edges, a graph is directed or undirected. Further, a graph is either weighted or unweighted. A sequence of nodes connected by (weighted) edges forms a (weighted) path. The shortest path is the one with the smallest sum over the weights \cite{b9}. A$^\star$ algorithm addresses the problem of finding the shortest path in a weighted graph\cite{b16}.

\begin{figure}
    \centering
    \includegraphics[width=0.5\textwidth]{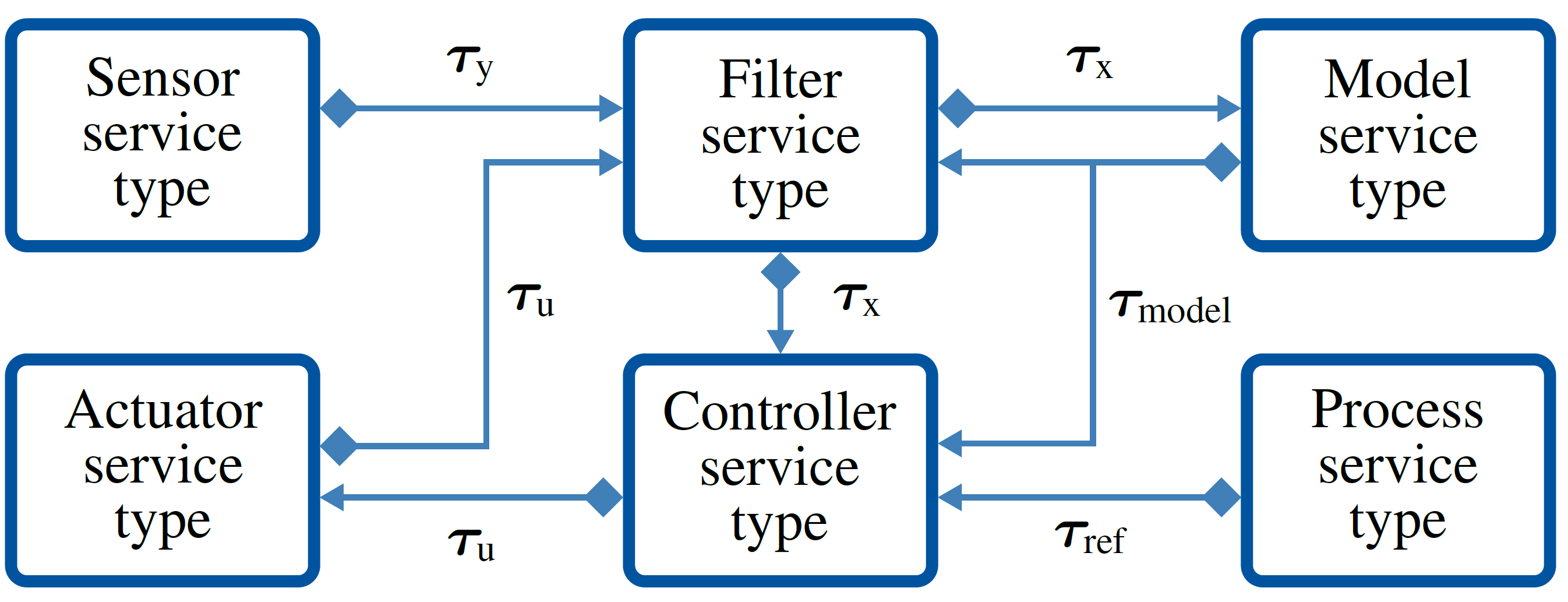}
    \caption{Control System Services.}
    \label{fig:basic_control_loop}
\end{figure}

In the context of \gls{SOA}, graphs are good representations  for service composition and orchestration, forming what is called a service graph. 
Figure~\ref{fig:basic_control_loop} presents a general architecture of a control system represented as a service graph. 
\begin{definition}[Service Graph]
A \emph{service graph} is a layered, directed graph 
$\mathcal{G}=(\mathcal{V},\mathcal{E},\ell,\mathcal{L})$ with:
\begin{itemize}
\item Vertex set $\mathcal{V}=\mathbb{S}\cup\{Start,End\}$, where $\mathbb{S}$ is the set of services.
\item Layers $\mathcal{L}=\{\mathcal{L}_1,\ldots,\mathcal{L}_K\}$ with a layer index 
$\ell:\mathcal{V}\to\{1,\ldots,K\}$ imposing the total order 
$\mathcal{L}_1\prec\cdots\prec\mathcal{L}_K$.
A canonical choice is 
$\mathcal{L}_1=\{Start\}$,
$\mathcal{L}_2=S_{\text{sensor}}$,
$\mathcal{L}_3=S_{\text{filter}}\cup S_{\text{model}}$,
$\mathcal{L}_4=S_{\text{controller}}\cup S_{\text{model}}$,
$\mathcal{L}_5=S_{\text{actuator}}$,
$\mathcal{L}_5=\{End\}$,
without any empty layer allowed.
\item Edge set $\mathcal{E}=\big\{(u,v)\in\mathcal{V}\times\mathcal{V}\;:\;\ell(u)<\ell(v)\ \text{ and }\ G_u \cap R_v \neq \emptyset.$

\end{itemize}
\end{definition}

\begin{definition}[Feasible Composition / Control-Loop Path]
A composition is a simple path 
$\pi=\langle s_0=\text{Start}, s_1,\ldots,s_M+1=\text{End}\rangle$ in $\mathcal{G}$ with strictly increasing layers, where $M$ is the number of services forming a control loop path.
It is feasible if for every internal node $s_j$ ($1\le j\le M$),
\[
R_{s_j}\ \subseteq\ \bigcup_{i=0}^{j-1} G_{s_i},
\]
i.e., all requirements of $s_j$ are provided by guarantees of some upstream services on the path.
\end{definition}

\begin{definition}[Weighted Service Graph]\label{def:weighted} A service graph where each edge has a weight defined by the cost function:
\[
w_\alpha(u\!\to\!v)\;=\;\alpha\,\tau(v)+(1-\alpha)\,\epsilon(v),
\quad (u\!\to\!v)\in\mathcal{E}.
\]
Each service $s\in\mathbb{S}$ has two nonnegative metrics: 
\begin{itemize}
\item Computation time $\tau(s)$: defined as the average execution time of each service
\item Inaccuracy $\epsilon(s)$: has different definitions according to the service type:
\begin{itemize}
\item Sensor: \gls{RMS} noise from datasheet
\item Model: \gls{RMS} of (model output - system output)
\item Filter: \gls{RMS} of (sensor RMS noise - abs(estimated value - measured values))
\item Controller: \gls{RMS} of reference tracking error
\item Actuator: datasheet accuracy/operating range
\end{itemize}
\end{itemize}
$\alpha\in[0.1,0.9]$ is a trade-off parameter. The range of this parameter was selected to avoid neglecting the effect of the inaccuracy or the computation time if $\alpha = 0$ or $\alpha = 1$. We could select the average or maximum for the inaccuracy instead of \gls{RMS}, but we preferred the \gls{RMS} following the convention of the \gls{MPC} cost function which optimizes the square of the error:
\[
J = 
\sum_{k=0}^{N-1}
(x_k - x_{\mathrm{ref}})^\top Q (x_k - x_{\mathrm{ref}})
+
(u_k - u_{\mathrm{ref}})^\top R (u_k - u_{\mathrm{ref}})
\]

\end{definition}
\begin{remark}[Accuracy vs Inaccuracy]
A$^\star$ algorithm finds paths with the lowest cost, so cost function variables must reflect that principle: lower values indicate better outcomes. To address this, we use inaccuracy (the inverse of accuracy).
\end{remark}

The aim of the service graph is to represent the selection cost of each service. To reach this goal, the graph is constructed similarly to the control system services in Figure~\ref{fig:basic_control_loop}, but with some differences:

\begin{enumerate} 
    \item We add a fixed start node and a fixed end node. The start node is connected to all available sensor nodes. All available actuator nodes are connected to the end node.\label{l}
    \item We remove the process service as we assume it is fixed and cannot be changed by the orchestrator. Note that the process service only provides the state reference and is not an abstraction for the real process or has a cost. 
    \item Controllers and filters may rely on a model. If a model is used, the model affects the controller or the filter cost, so filters/controllers and the respective models
are grouped into a single vertex.
    \item We remove the edge from the actuator to the filter as we already considered the cost of the filter selection from the edge between the sensor and the filter, so it will be redundant.
\end{enumerate}

\begin{figure}
  \centering
  \includegraphics[width=0.5\textwidth]{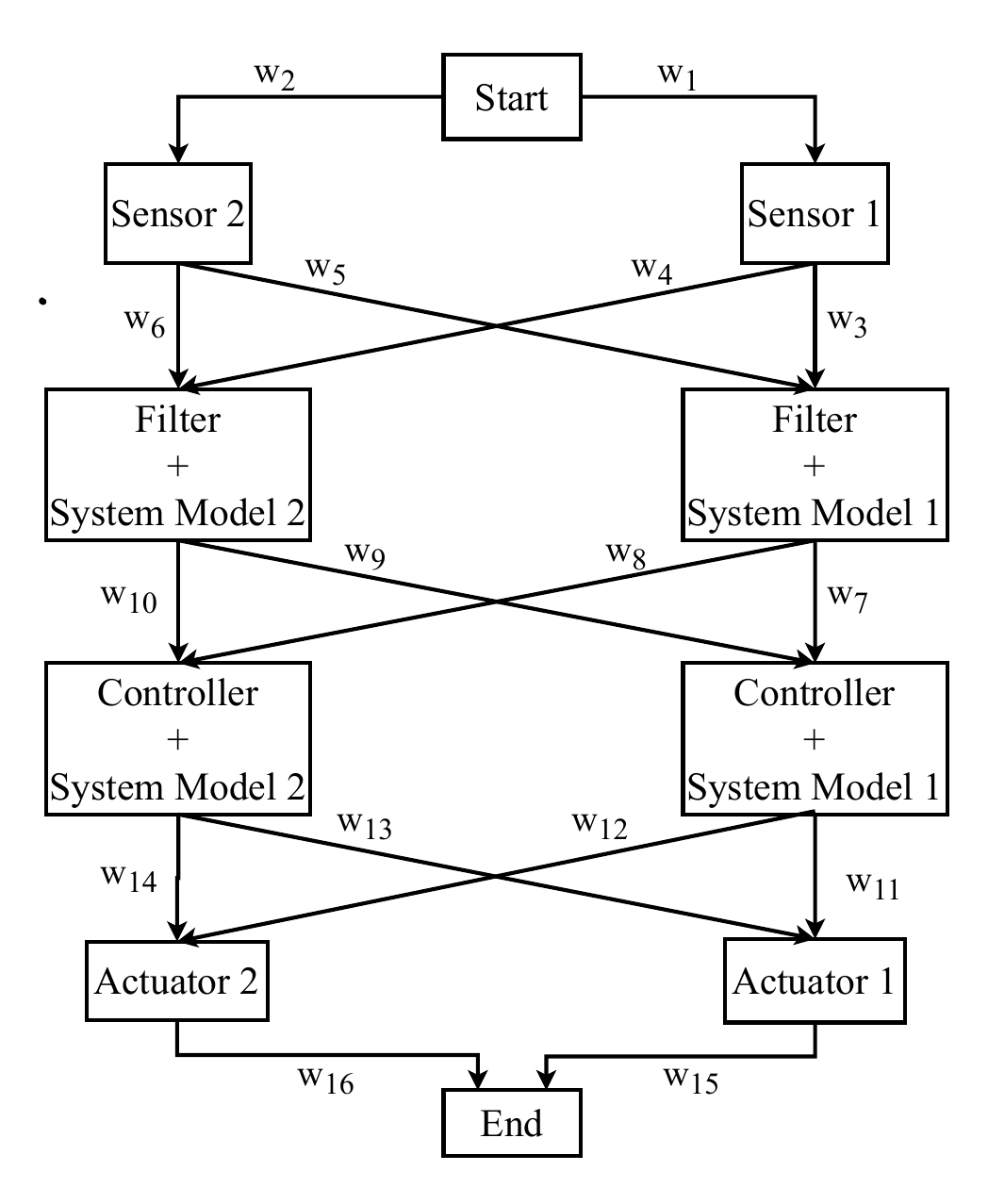}
  \caption{Service Graph.}
  \label{fig:starttarget}
\end{figure}

Applying these steps to the control system services example in Figure \ref{fig:basic_control_loop} and assuming two available services for sensors, models, and actuators results in the service graph in Figure \ref{fig:starttarget}. 

The steps presented in 1)-4) illustrate how we transform the control system services into a service graph to be ready for the A$^\star$  orchestration. We simplify this process by directly building the graph from the list of available services. Algorithm \ref{alg:create_graph} outlines the procedure.

\begin{algorithm}
\caption{CreateServiceGraph}
\label{alg:create_graph}
\begin{algorithmic}[1]
\Require $S_{\text{sensor}}$, $S_{\text{filter}}$, $S_{\text{model}}$, $S_{\text{controller}}$, $S_{\text{actuator}}$, and cost function $w_\alpha$
\Ensure service graph $G$

\State $G \gets \emptyset$
\State Add nodes $\textit{Start}, S_{\text{sensor}}$ to $G$
\State Connect \textit{Start} to $S_{\text{sensor}}$

\For{each $s_{\text{filter}}  \in S_{\text{filter}}$}
    \If{$s_{\text{filter}} $ requires a model}
        \For{each $s_{\text{model}} \in S_{\text{model}}$}
            \State Add $(s_{\text{filter}},s_{\text{model}})$ to $G$; connect them to $S_{\text{sensor}}$ 
        \EndFor
    \Else
        \State Add $s_{\text{filter}}$ to $G$; connect it to $S_{\text{sensor}}$
    \EndIf
\EndFor

\For{each $s_{\text{controller}}\in S_{\text{controller}}$}
    \If{$s_{\text{controller}}$ requires a model}
        \For{each $s_{\text{model}} \in S_{\text{model}}$}
            \State Add $(s_{\text{controller}},s_{\text{model}})$ to $G$
        \EndFor
    \Else
        \State Add $s_{\text{controller}}$ to $G$
    \EndIf
    \For{each $s_{\text{filter}}$ or $(s_{\text{filter}},s_{\text{model}})$}
            \State Connect it to $s_{\text{controller}}$ or $(s_{\text{controller}},s_{\text{model}})$
    \EndFor
\EndFor

\State Add $S_{\text{actuator}}$; connect them to all $s_{\text{controller}}$ or $(s_{\text{controller}},s_{\text{model}})$
\State Add \textit{End} node; connect $S_{\text{actuator}}$ to it
\State Add edge weights according to $w_\alpha$
\State \Return $G$

\end{algorithmic}
\end{algorithm}
First, we initialize the service graph (line 1). Then we add the start node and sensors services (line 2). After that, we connect every sensor service to the start node (line 3). Next, we add filters services (lines 4-12). If the filter requires a model, we add every possible combination of the filter with the available models as separate nodes. We connect all the filters to all the sensors. For the controllers, we do the same (lines 13-24). We add the actuator services and connect them to the controller services (line 25). We add the end node and connect it to every actuator service (line 26). Lastly, we compute the cost of choosing each service and add it as a weight to all its incoming edges (line 27).

\subsection{A$^\star$ Orchestration}
\label{sect:orchworkflow}
\begin{figure}
  \centering
  \includegraphics[width=0.5\textwidth]{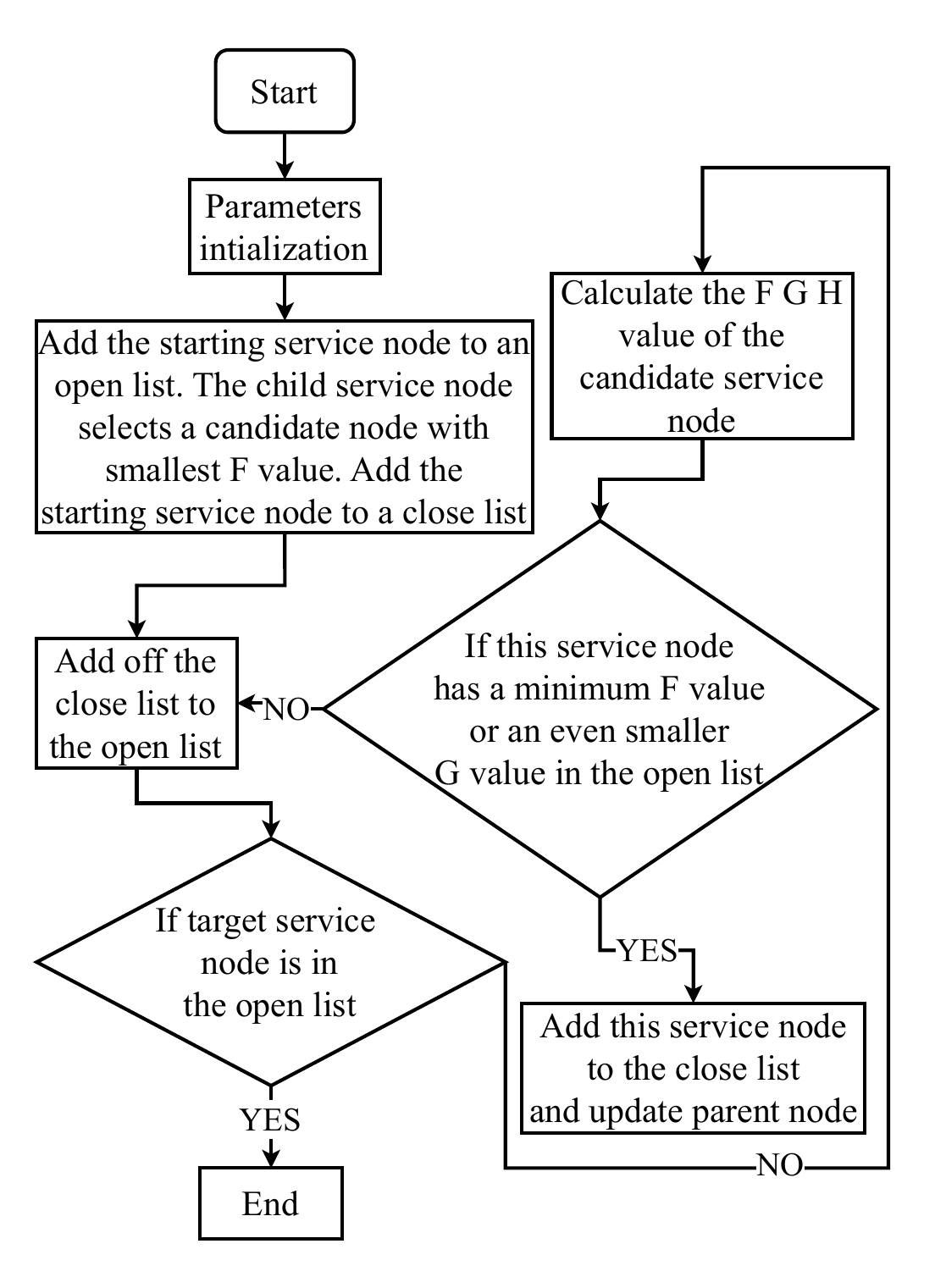}
  \caption{A$^\star$ Orchestration\cite{b16}.}
  \label{fig:A*}
\end{figure}

\begin{algorithm}
\caption{Orchestrate}
\label{alg:orchestrator_workflow}
\begin{algorithmic}[1]
\Require services, cost function $w_\alpha$, evaluation criteria $F(\alpha)$
\State \textit{$\alpha^\star$} $\gets$ Contextual \gls{BO}($F(\alpha)$)
\State \textit{service graph} $\gets$ CreateGraph(services, $w_\alpha$)
\State \textit{optimalPath} $\gets$ A$^\star$(service graph, start, target)
\State set \textit{optimalPath} as new control system architecture

\end{algorithmic}
\end{algorithm}
After building the service graph, the orchestrator uses A$^\star$ path planning algorithm to find the optimal feasible service composition.
\begin{definition}[Optimal Service Composition]
The cost of a composition $\pi$ is defined as
$J(\pi;\alpha)=\sum_{(u\to v)\in \pi} w_\alpha(u\to v)$. So the optimal service composition is defined as the optimal path:
\[
\pi^\star(\alpha)\in\arg\min_{\pi\ \text{feasible}} J(\pi;\alpha).
\]
\end{definition}
The A$^\star$ algorithm uses two important parameters to find the cost of a path:
\begin{itemize}
    \item $g(n)$: Actual cost to reach the node $n$ from the start node.
    \item $h(n)$: The heuristic finds the cost of reaching the goal from node $n$, for which we used a greedy selection of the minimum subsequent service composition to reach the goal \[
w_\alpha^\star(u\!\to\!v)
\]
\end{itemize}
The function $f(n)=g(n)+h(n)$ is the total estimated cost of the optimal solution through node $n$ and is used in the algorithm workflow as shown in the algorithm in Figure \ref{fig:A*}.

\begin{figure}
  \centering
  \includegraphics[width=0.5\textwidth]{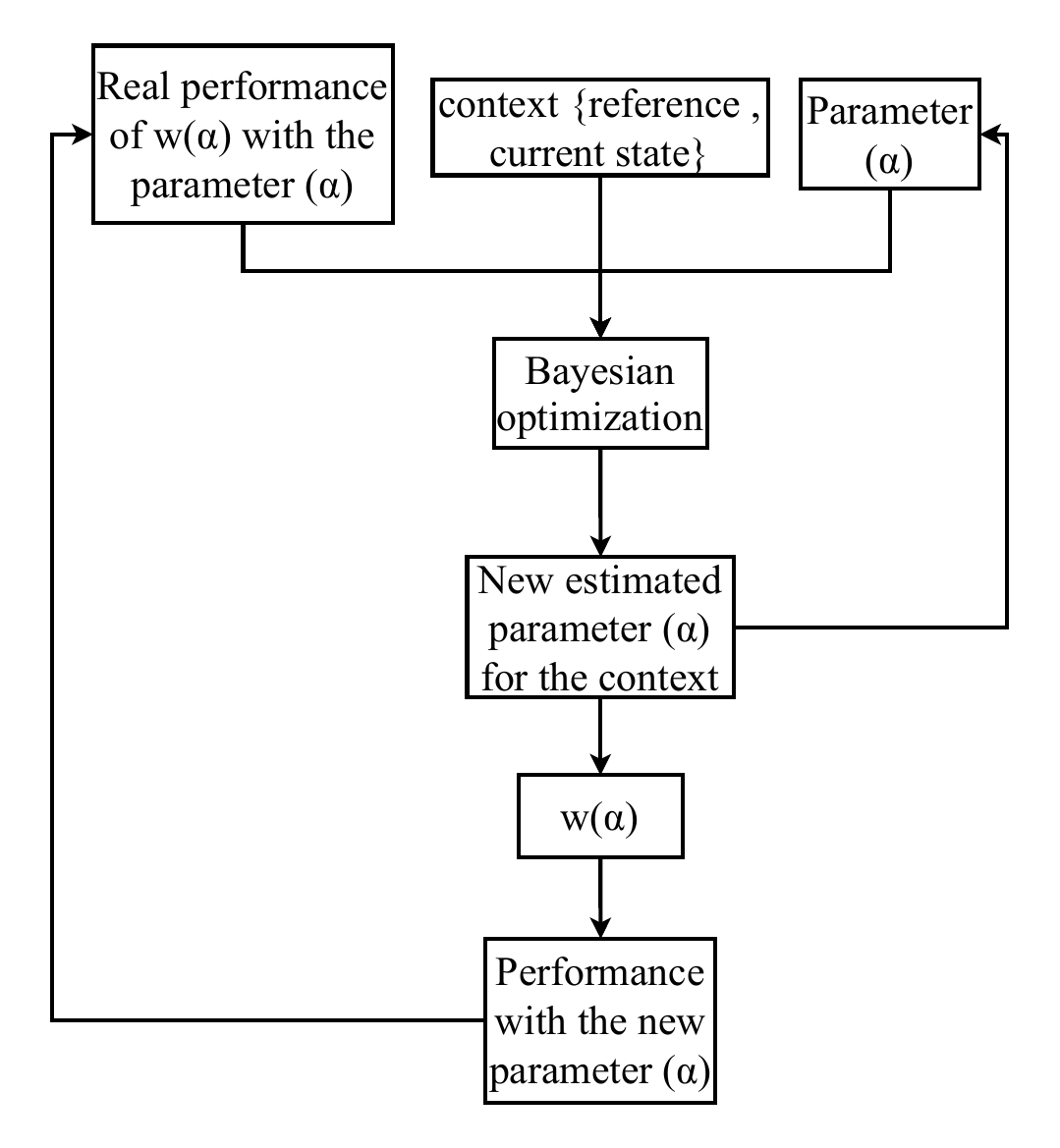}
  \caption{Contextual \gls{BO} for trade-off $\alpha$ parameter estimation.}
  \label{fig:opt}
\end{figure}

The last part of our orchestration is the selection of the trade-off parameter $\alpha$ to achieve the trade-off between the computation time and the inaccuracy of the optimal service composition $\pi^\star(\alpha)$. Contextual \gls{BO} \cite{b12} over $[0.1,0.9]$ is used to estimate this parameter based on the current context, which is defined in terms of the reference value and the current state, as shown in the algorithm in Figure \ref{fig:opt}, such that:
\[
\alpha^\star \in \arg\min_{\alpha\in[0.1,0.9]} F(\alpha),
\]
where $F(\alpha)$ is the preferred evaluation criterion, which could be selected to decrease the inaccuracy or the computation time.

Given the method for creating the service graph, we can now outline the system orchestration process in Algorithm \ref{alg:orchestrator_workflow}. The first step is to estimate the trade-off parameter $\alpha$ using Contextual \gls{BO}. The next step is the creation of the service graph by conducting Algorithm \ref{alg:create_graph}. Once this graph is compiled, the orchestrator determines the optimal service composition by applying the A$^\star$ algorithm. The resulting path is then set as the control system architecture. This process repeats every time the cost function changes or a service is removed, updated, or added.

\section{Evaluation}
\label{sect:evaluation}
To evaluate our framework, we used a vehicle longitudinal velocity control case study subjected to maximum vehicle velocity and acceleration. We used three vehicle models with different dynamics complexity \cite{b26}:
\begin{itemize}
 \item Point-Mass Model.
 \item Single-Track Model.
 \item Multi-Body Model.
\end{itemize}
We also considered the multi-body model to be the one with zero inaccuracy, so other models outputs' could be compared to it according to Definition \ref{def:weighted}. For these evaluation scenarios, our control system consists of:
\begin{itemize}
\item Three models 
\item Two sensors 
\item Kalman filter
\item Dummy filter (which just passes the measured states without filtering)
 \item Two actuators
 \item Two controllers for tracking reference longitudinal velocity: 
\begin{itemize}
 \item PID Controller
 \item \gls{MPC}
\end{itemize}
\end{itemize}
This case study was used to assess two evaluation scenarios: (i) reference tracking error, (ii) computation time as the evaluation criterion of the contextual \gls{BO} during runtime. The first step of the evaluation is to give initial estimates of the inaccuracy and computation time for each service. Then, based on the evaluation criterion, the contextual \gls{BO} gives an initial estimate of the trade-off parameter $\alpha$. After that, a service graph is created from the available services, and the orchestrator discovers the optimal service composition. The system continues to run and update the inaccuracy $\epsilon(s)$ and the computation time $\tau(s)$ of each service according to Definition \ref{def:weighted}. The contextual \gls{BO} also continues to learn and estimate the optimal trade-off parameter $\alpha^\star$ for each context. We selected an initial longitudinal velocity of 1 m/s and a reference of 6 m/s. We evaluated the two scenarios for this context.
\begin{figure*}[h!]
    \centering
    \begin{subfigure}{\textwidth}
        \centering
        \includegraphics[width=\linewidth]{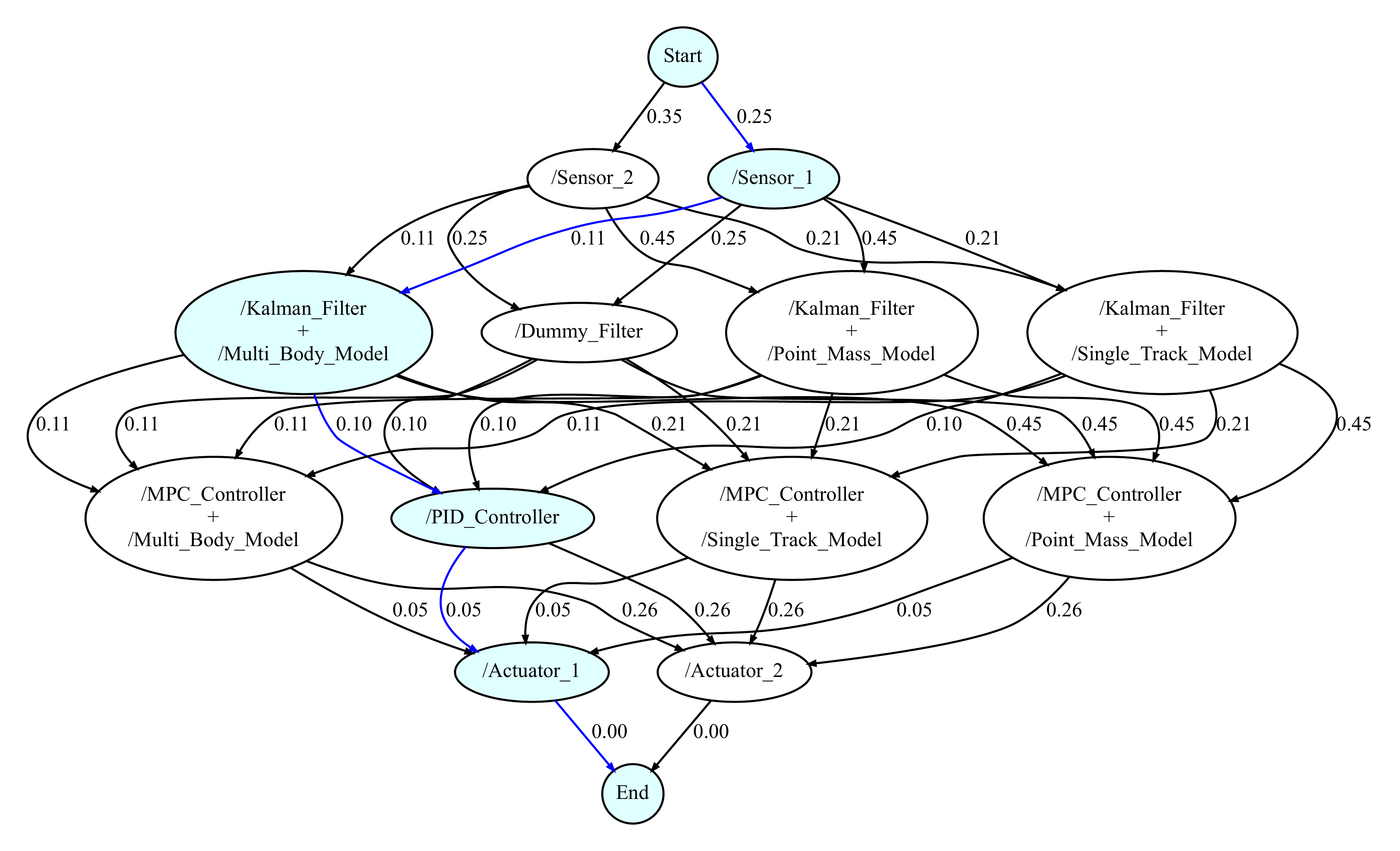}
        \caption{Optimal service composition for $\alpha = 0.5$ and context of $(1,6)$.}
        \label{fig:first_e}
    \end{subfigure}
    \hfill
    \begin{subfigure}{\textwidth}
        \centering
        \includegraphics[width=\linewidth]{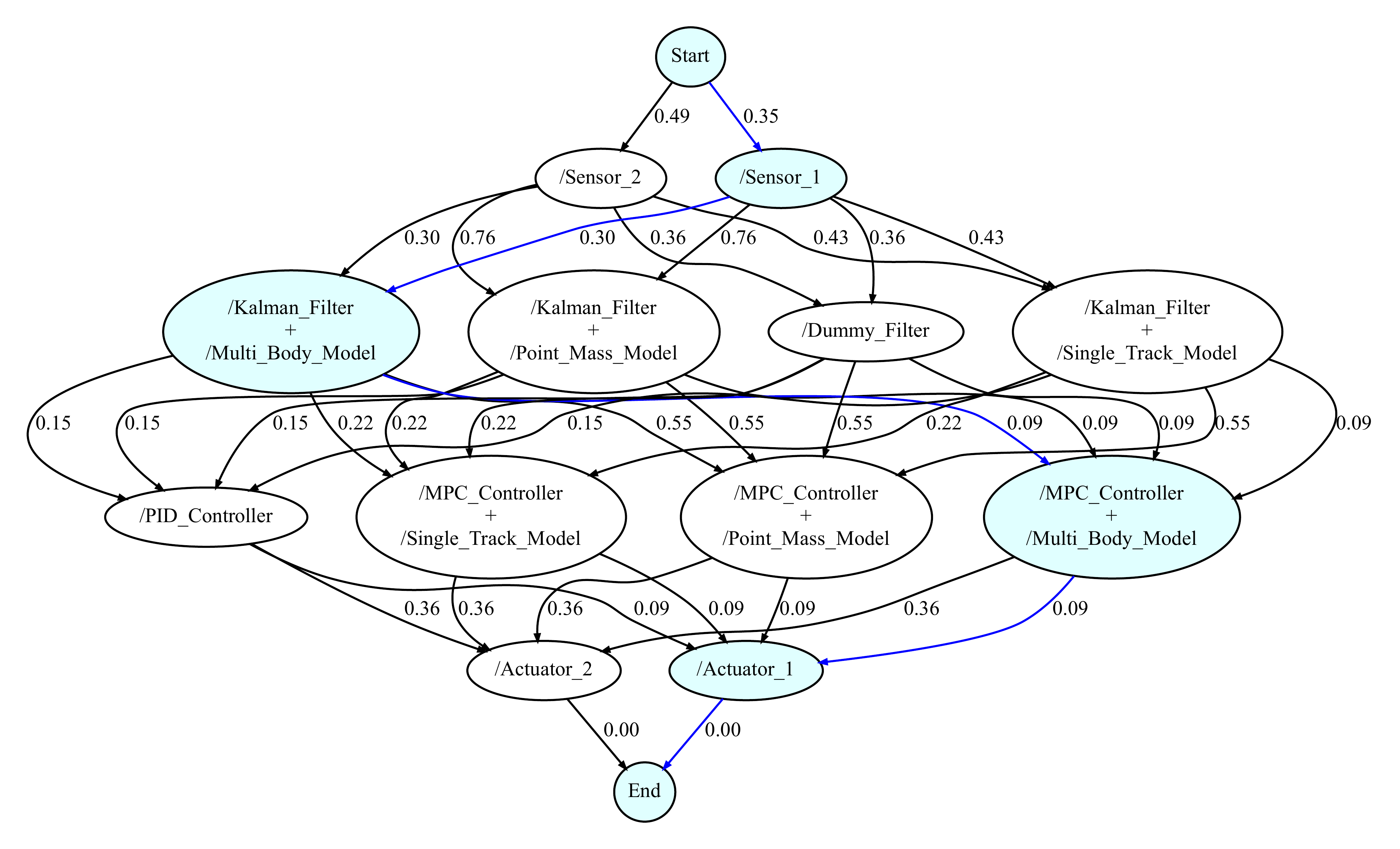}
        \caption{Reference tracking performance for $\alpha = 0.1$ and context of $(1,6)$.}
        \label{fig:final_e}
    \end{subfigure}
    \caption{Initial and final optimal service composition in (a) and (b) respectively before and after converging to the optimal trade-off parameter $\alpha^\star$ for reference tracking error optimization}
\end{figure*}
\begin{figure*}[h!]
    \centering
    \begin{subfigure}{0.4978\textwidth}
        \centering
        \includegraphics[width=\linewidth]{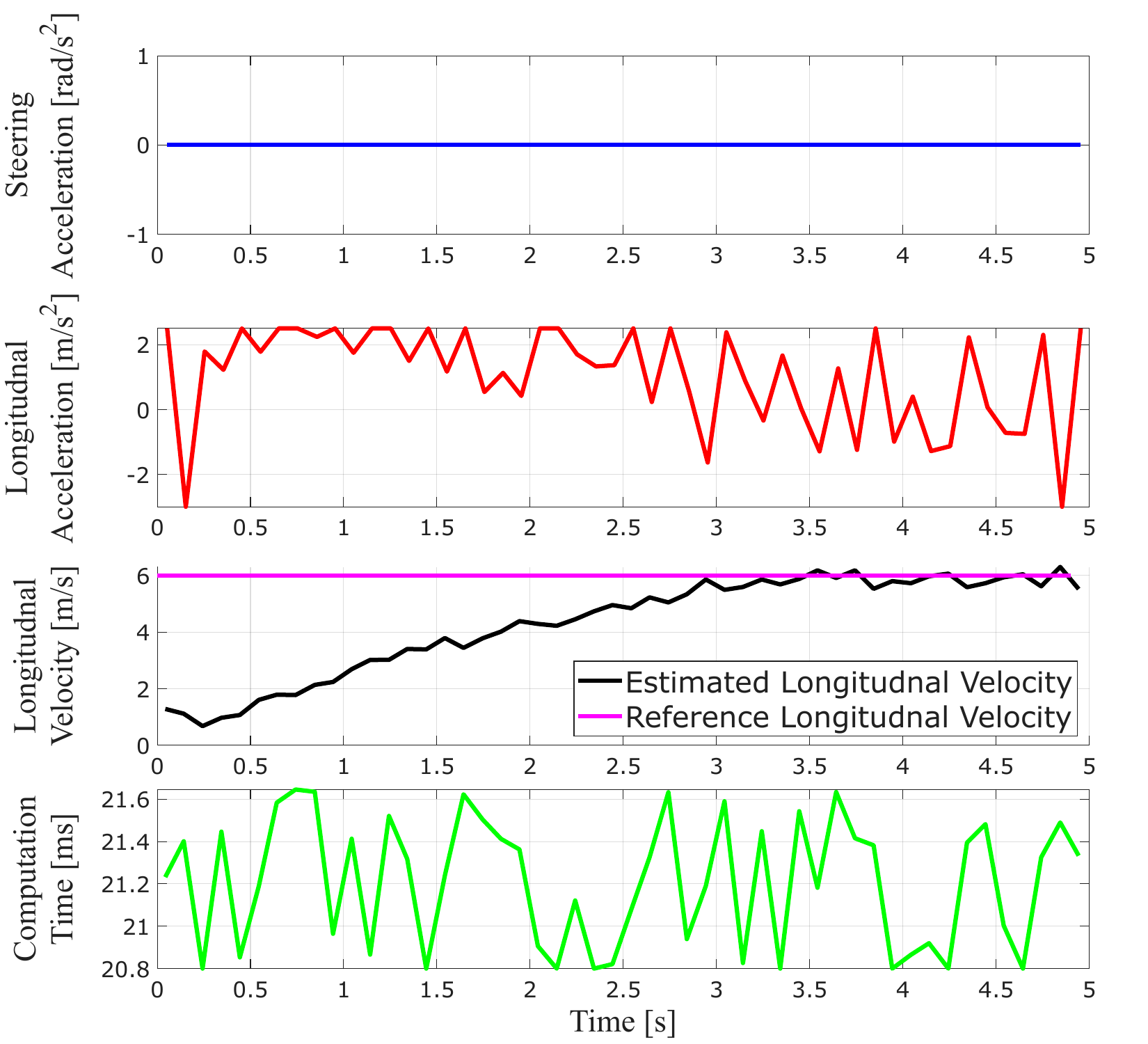}
        \caption{Reference tracking performance for $\alpha = 0.5$ and context of $(1,6)$.}
        \label{fig:first_er}
    \end{subfigure}
    \hfill
    \begin{subfigure}{0.4978\textwidth}
        \centering
        \includegraphics[width=\linewidth]{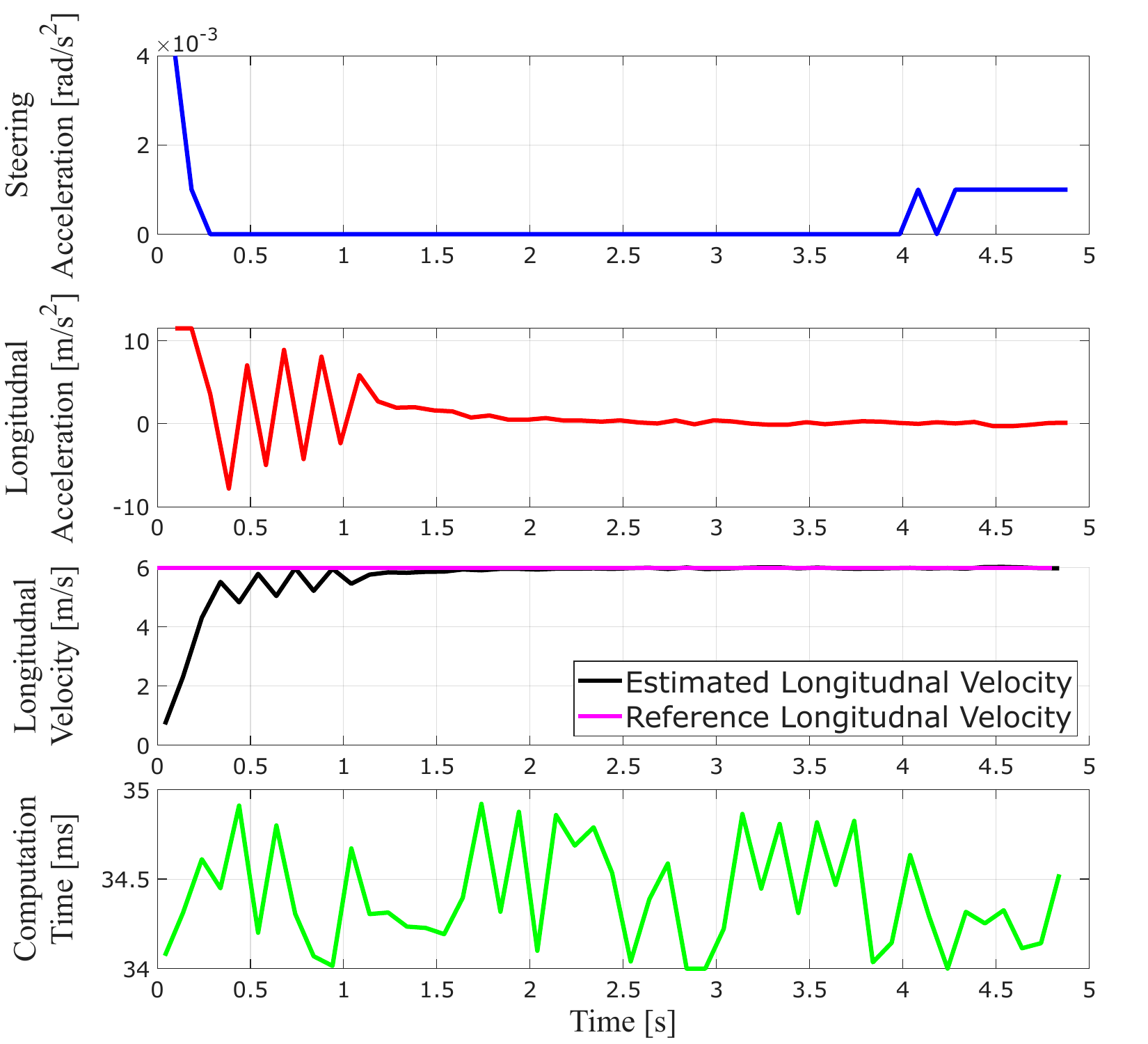}
        \caption{Reference tracking performance for $\alpha = 0.1$ and context of $(1,6)$.}
        \label{fig:final_er}
    \end{subfigure}
    \caption{Initial and final reference tracking performance in (a) and (b), respectively, before and after converging to the optimal trade-off parameter $\alpha^\star$ for reference tracking error optimization.}
\end{figure*}
\begin{figure*}[h!]
    \centering
    \begin{subfigure}{0.4978\textwidth}
      \centering
      \includegraphics[width=\textwidth]{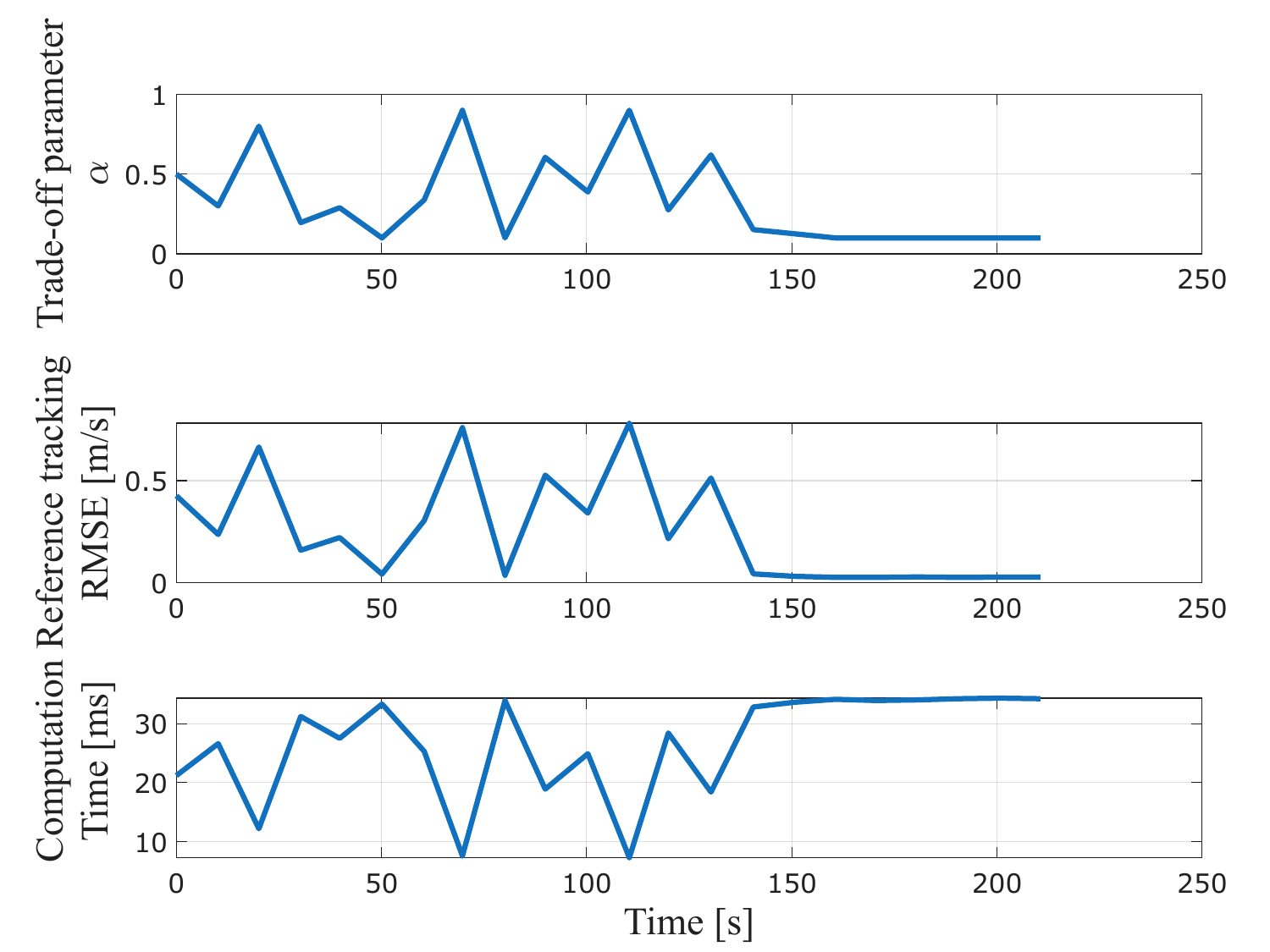}
      \caption{Reference tracking error optimization at context of (1, 6).}
      \label{fig:eval_e}
    \end{subfigure}
    \hfill
    \begin{subfigure}{0.4978\textwidth}
      \centering
      \includegraphics[width=\textwidth]{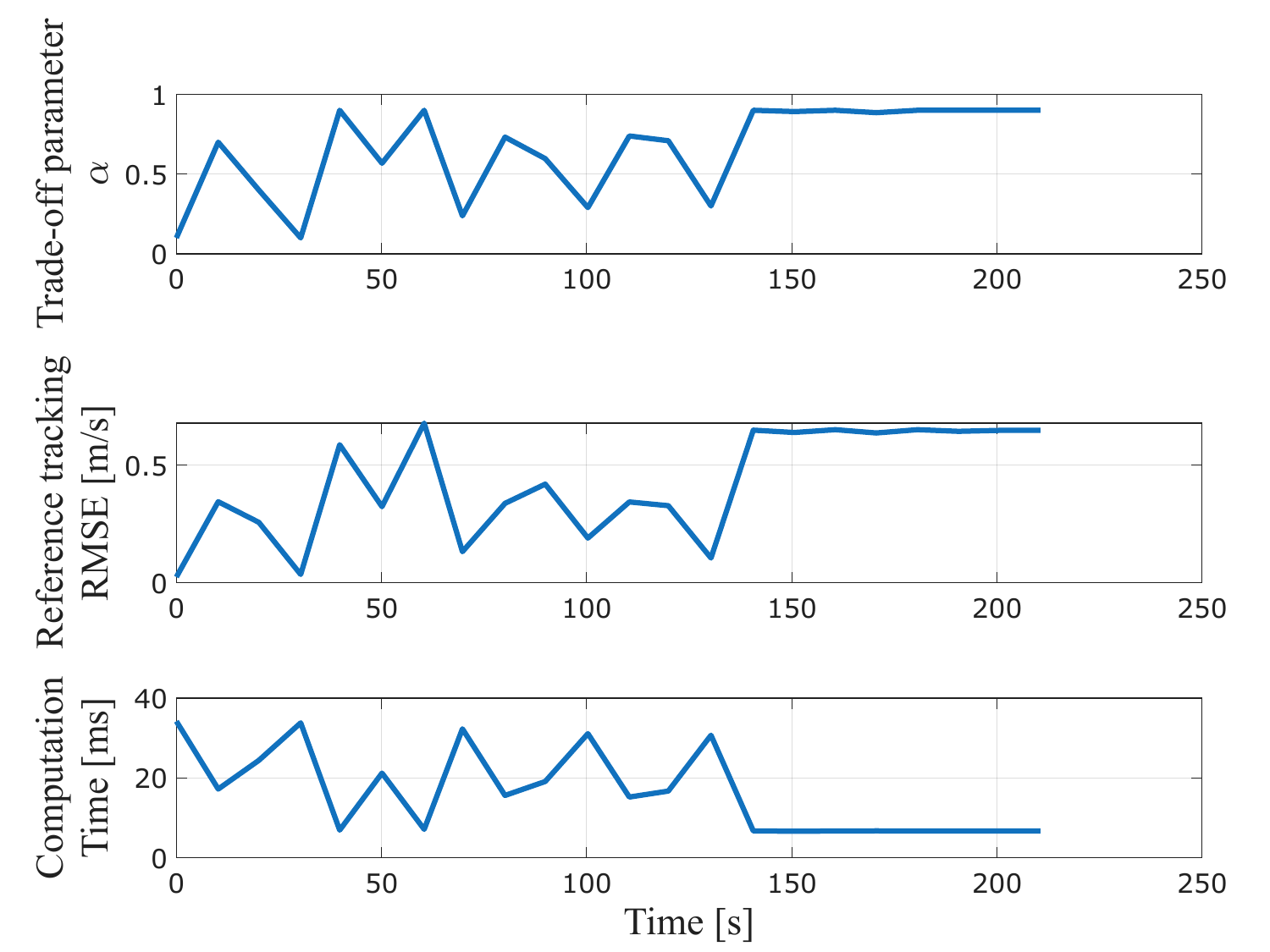}
      \caption{Computation time optimization at context of (1, 6).}
      \label{fig:eval_c}
    \end{subfigure}
    \caption{Contextual \gls{BO} learning process.}
\end{figure*}
\begin{figure*}
  \centering
  \includegraphics[width=\textwidth]{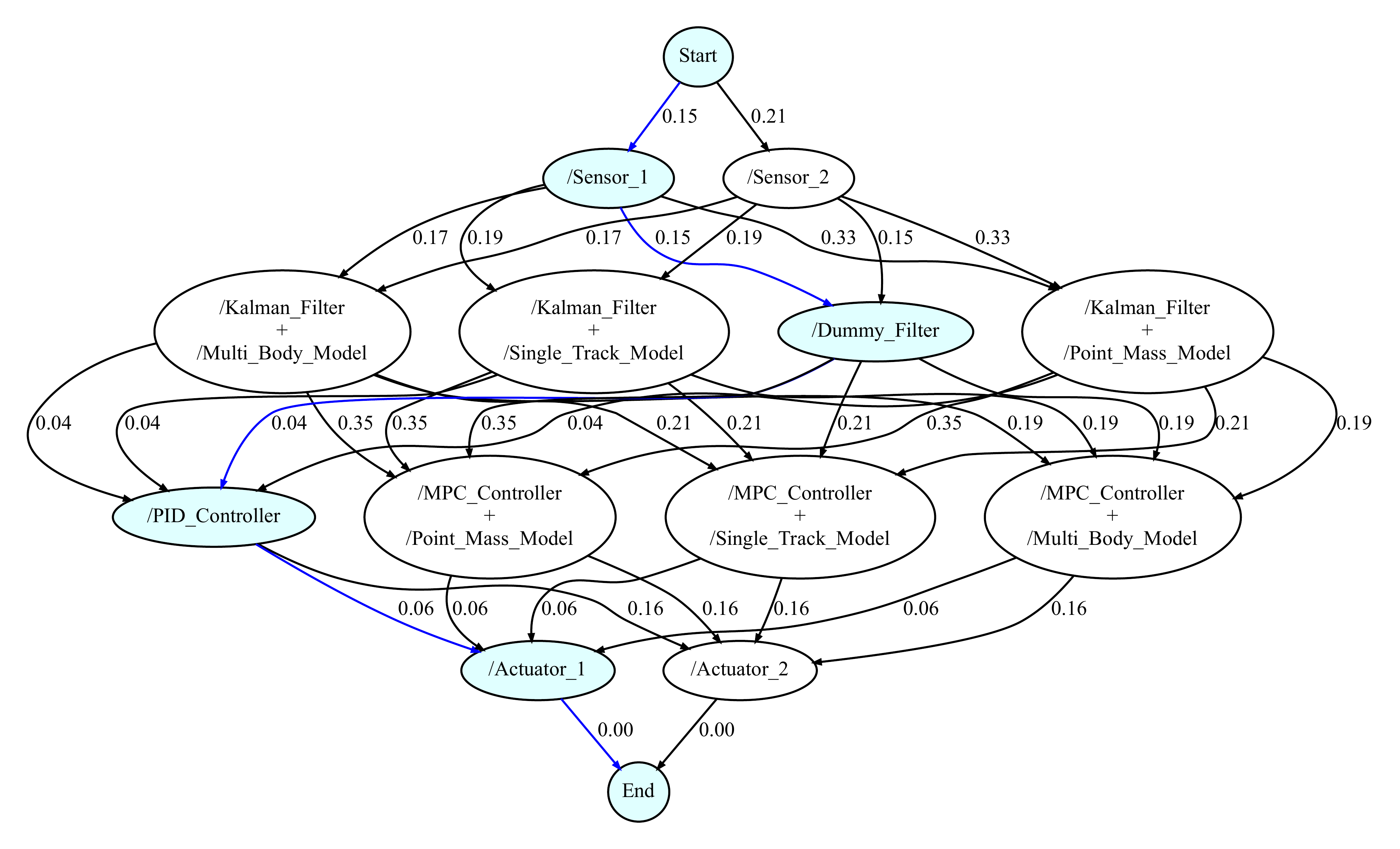}
  \caption{Final optimal service composition after converging to the optimal trade-off parameter $\alpha^\star = 0.9$ for computation time optimization at context of (1, 6).}
  \label{fig:final_c}
\end{figure*}
\begin{figure}
  \centering
  \includegraphics[width=0.5\textwidth]{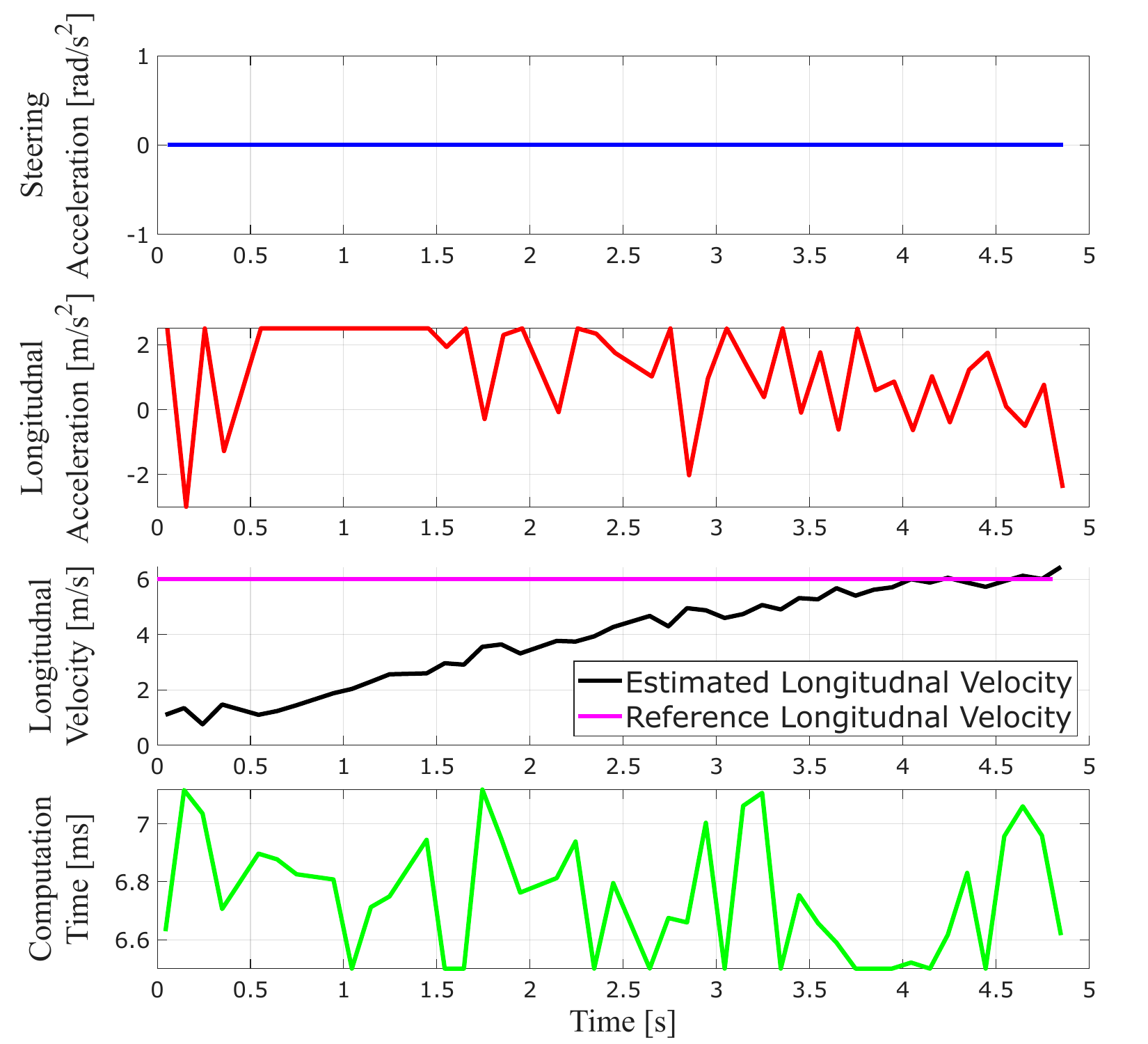}
  \caption{Final reference tracking and computation time performance after converging to the optimal trade-off parameter $\alpha^\star = 0.9$ for computation time optimization at context of (1, 6).}
  \label{fig:final_cr}
\end{figure}
\begin{table}[!htb]
    \centering
    \caption{Optimal Service Composition during Contextual \gls{BO} learning}
    \begin{tabular}{|c|p{1.5cm}|p{1.7cm}|p{0.7cm}|p{1.5cm}|}
        \hline
        \textbf{$\alpha$} & \textbf{Controller} & \textbf{Filter} & \textbf{RMSE [m/s]}& \textbf{Computation time [ms]}  \\
        \hline
        0.732 & PID & Kalman+Single Track Model & 0.338 & 15.6\\ \hline
        0.4 & MPC+Single Track Model & Kalman+Single Track Mode & 0.257 & 24.4\\ \hline
        0.29 & MPC+Multi Body Model & Kalman+Point Mass Mode & 0.19 & 29.5 \\ \hline
        0.8 & PID & Kalman+Point Mass Mode & 0.665 & 12.2 \\ \hline
        0.196 & MPC+Multi Body Model & Kalman+Single Track Mode & 0.15 & 31.2 \\ \hline
    \end{tabular}
    \label{tab:first}
\end{table}

For the first evaluation scenario, the contextual \gls{BO} started with an estimate of $\alpha = 0.5$. This value represents a balance between computation time and inaccuracy optimization. Our framework was adapted to this value and selected the optimal service composition shown in Figure \ref{fig:first_e}, confirming this balance by choosing the PID controller, which has low computation time, and the Kalman filter, which has high accuracy. The selected composition achieved a tracking performance shown in Figure \ref{fig:first_er} with an \gls{RMS} tracking error of 0.4 m/s and an average computation time of 21.2 ms. The contextual \gls{BO} kept exploring and exploiting different values of $\alpha$, leading to different service compositions; some of them are listed in Table \ref{tab:first}, knowing that $Sensor\_1$ and $Actuator\_1$ were selected in all compositions. The selected models in Table \ref{tab:first} behaved as expected, as the most complex model (Multi-Body Model) showed the highest accuracy and computation time, while the least complex model (Point-Mass Model) resulted in the lowest accuracy and computation time. The contextual \gls{BO} finally converged to $\alpha = 0.1$, as shown in Figure \ref{fig:eval_e}. This value yielded the optimal service composition in Figure \ref{fig:final_e}, with a reduced \gls{RMS} tracking error of 0.026 m/s and an increased computation time to 34.4 ms, as shown in Figure \ref{fig:final_er}. This optimal tracking performance resulted from the high accuracy of the \gls{MPC} and Kalman filter with the multi-body model.

After some time, the evaluation criterion changed automatically at runtime to prefer computation time optimization. The contextual \gls{BO} continues from the last estimated value of $\alpha = 0.1$. This value indicates a preference for the inaccuracy optimization. Based on this preference, our framework selected the optimal service composition shown in Figure \ref{fig:final_e}, which is the same as that chosen after optimizing the reference tracking error in the first scenario, with the same tracking performance shown in Figure \ref{fig:final_er}. The contextual \gls{BO} performed some exploration and exploitation until it converged to $\alpha = 0.9$, as shown in Figure \ref{fig:eval_c}. The optimal service composition resulting from this value showed a strong preference for low computation time, as indicated by the selection of the PID controller and the Dummy filter in Figure \ref{fig:final_c}. This composition achieved a reduced computation time of 6.7 ms and a high \gls{RMS} tracking error of 0.69 m/s, as shown in Figure \ref{fig:final_cr}.

\section{Conclusion and Future Work}
\label{sect:conclusion}
This work introduces a framework for computation-accuracy trade-off in \gls{SOMC}. Our framework significantly improves computation time or inaccuracy based on a preferred evaluation criterion. This improvement is due to the ability to perform dynamic changes at runtime in the control system until reaching the optimal service composition. The evaluation shows how our framework succeeded in reaching an optimal service composition in case of inaccuracy or computation time optimization.

Future work includes extending the cost function by adding a latency variable to account for data transmission times between services. Additionally, we will define the requirements for integrating an AI service to support performance-related decisions. Moreover, we plan to test machine learning and reinforcement learning control algorithms.

\section*{Acknowledgment}
The authors wish to express their sincere thanks to  Jonathan Kelm for the valuable discussions.

\end{document}